\documentclass[letterpaper, 10 pt, conference]{ieeeconf}  

\usepackage{amsmath,amssymb,amsfonts}
\usepackage{graphicx,wrapfig}
\usepackage{subcaption}
\usepackage{makecell}

\usepackage{gensymb}
\usepackage{color}
\usepackage[colorlinks]{hyperref}

\usepackage{accents}
\newcommand{\ubar}[1]{\underaccent{\bar}{#1}}

\def\rd#1{{\color{red}{#1}}}
\def\pb#1{\footnote{PB:\rd{#1}}}

\IEEEoverridecommandlockouts                              

\usepackage{xparse}

\ExplSyntaxOn
\cs_new:Nn \pb_prop_gset_bykeys:Nn
{
	\prop_clear:N \l__pb_temp_prop
	\keys_set:nn { pb/propbykey } { #2 }
	\prop_gset_eq:NN #1 \l__pb_temp_prop
}
\cs_generate_variant:Nn \pb_prop_gset_bykeys:Nn { c }

\keys_define:nn { pb/propbykey }
{
	unknown .code:n = \prop_put:Nxn \l__pb_temp_prop { \l_keys_key_tl } { #1 }
}

\cs_generate_variant:Nn \prop_put:Nnn { Nx }

\NewDocumentCommand{\setupcollaborator}{mm}
{
	\prop_new:c { g_collaborator_#1_prop }
	\pb_prop_gset_bykeys:cn { g_collaborator_#1_prop } { #2 }
}

\NewDocumentCommand{\selectcollaborator}{m}
{
	\prop_map_inline:cn { g_collaborator_#1_prop }
	{
		\tl_set:cn { ##1 } { ##2 }
	}
}
\ExplSyntaxOff

\setupcollaborator{PBmac}
{
	NAME=Prabir Mac , laptop or mac mini,
	DiCEbibPATH=/Users/pbarooah/Dropbox/DiCElab_bib,
	NRbibPATH = /Users/pbarooah/Dropbox/NSRbib
}
\setupcollaborator{NRlinux}
{
	NAME=NR at lab,
	DiCEbibPATH=../../DiCElab_bib,
	NRbibPATH = ../../NSRbib,
}
\setupcollaborator{NRwindows}
{
	NAME=NR using windows,
	DiCEbibPATH=../../DiCElab_bib,
	NRbibPATH = ../../NSRbib,
}
\setupcollaborator{NGlinux}
{
	NAME=NG using windows,
	DiCEbibPATH=../../../../../DiCElab_bib,
}

\selectcollaborator{NGlinux} 

\graphicspath{{./Images/}}

\title{\LARGE \bf
Smart Home Energy Management System for Power System Resiliency
}

\author{Ninad Gaikwad, Naren Srivaths Raman, and Prabir Barooah
\thanks{The authors are with the Department of Mechanical and Aerospace Engineering, University of Florida, Gainesville, Florida 32611, USA.
	{\tt\small ninadgaikwad@ufl.edu, narensraman@ufl.edu, pbarooah@ufl.edu.} The research reported here is partially supported by NSF awards 1646229 and 1934322.}%
}

\begin{document}

\maketitle
\thispagestyle{empty}
\pagestyle{empty}

\begin{abstract}
	
The need for resiliency of electricity supply is increasing due to increasing frequency of natural disasters---such as hurricanes---that disrupt supply from the power grid. Rooftop solar photovoltaic (PV) panels together with batteries can provide resiliency in many scenarios. Without intelligent and automated decision making that can trade off conflicting requirements,  a large PV system and a large battery is needed to provide meaningful resiliency. By using forecast of solar generation and household demand, an intelligent decision maker can operate the equipment (battery and critical loads) to ensure that the critical loads are serviced to the maximum duration possible. With the aid of such an intelligent control system, a smaller (and thus lower cost) system can service the primary loads for the same duration that a much larger system will be needed to service otherwise.

In this paper we propose such an intelligent control system. A model predictive control (MPC) architecture is used that uses available measurements and forecasts to make optimal decisions for batteries and critical loads in real time. The optimization problem is formulated as a MILP (mixed integer linear program) due to the on/off decisions for the loads. Performance is compared with a non-intelligent baseline controller, for a PV-battery system chosen carefully for a single family house in Florida. Simulations are conducted for a one week period during hurricane Irma in 2017. Simulations show that the cost of the PV+battery system to provide a certain resiliency performance, duration the primary load can be serviced successfully, can be halved by the proposed control system. 
\end{abstract}

\section{Introduction}\label{section:Intro}

Extreme climate events are becoming more common the world over. 
In the United States, hurricanes, heat waves and forest fires are occurring with increasing frequency~\cite{USGCRPreport:2014}. Among the many consequences of these natural disasters, one is the loss of electricity supply for long periods. A few recent examples include 4.8 millions of utility customers losing electricity in Florida after hurricane Irma, with 1.5 million remaining without electricity for five days or more~\cite{IrmaImpactReport:EIA:2017}, and
the months-long blackout in Puerto Rico after hurricane Maria~\cite{GallucciRebuilding:spectrum:2018}, leading to an estimated death toll in the thousands~\cite{KishoreMortality:NEJM:2018}.

Distributed solar generation can provide a resilient energy supply since  the sky is often clear  immediately after the hurricane. However, as the average household load in the U.S. is quite high $30.5~kWh/day$~\cite{EIAAnnual:2019}, serving the entire household load from an on-site PV+battery system will require a large system, driving up cost substantially.

We argue that the size---and thus, cost---of the PV+storage system to provide resiliency can be reduced with the help of an intelligent control system. The key is to exploit flexibility in the demand as well as in the supply in conjunction with forecasts. Flexibility in demand comes from the fact that, after a disaster not all households loads need to be served. Among the critical loads that are needed to be served, refrigeration for food and medicine is the most important~\cite{CdcKeep:2019}. Next comes lights, and then fans. Fans can serve as temporary replacements for air conditioners to provide thermal comfort, and are much less energy intensive than  air conditioners. An intelligent controller can prioritize the refrigerator demand over light and fan demand, and turn off all other loads. Flexibility in supply comes from the fact that the charging rate of batteries is variable; a battery can be fast charged to prepare for a forecasted low solar irradiance event, though at some cost to the  battery's health.

Thus, by using forecast of solar generation and household demand, an intelligent decision maker can operate the equipment (battery, primary loads and secondary loads) to ensure that the critical loads are serviced to the maximum duration possible. With the aid of such an intelligent control system, a smaller (and thus lower cost) system can service the critical loads for the same duration that a much larger system will be needed to service otherwise.

In this paper we propose such an intelligent control system, for a home with three critical loads (a refrigerator, a few lights and fans) and a small PV+battery system. Among the three critical loads that need to be served with on-site energy after an outage, it is more important to serve the refrigerator than the lights and fans. We call the refrigerator the \emph{primary} load and the aggregate of the lights and fans the \emph{secondary} load.  Although there are many more electrical loads in a typical household, others are not critical for health and well being after a disaster. For instance, though many homes use electric cook-tops for cooking in Florida, people often use outdoor gas grills to cook food after hurricanes~\cite{NorrisHow:2017,OttensteinGet:2019}. The goal of the control system is to keep the refrigerator temperature within a band while serving the secondary demand as much as possible. 

A model predictive control (MPC) architecture is used that uses available measurements and forecasts to make optimal decisions in real time. The optimization problem is formulated as a MILP (mixed integer linear program); the integer valued variables are for the on/off status of the two loads, the primary load (refrigerator) and the aggregate demand of lights and fans. A dynamic model of the refrigerator is used to decide its compressor on/off status so that its temperature stays within an allowable band. Since fast charging of the battery degrades life, the controller tries to use normal charging as much as possible, using fast charging only when absolutely necessary.

In this preliminary work we only focus on the control algorithm for post-disaster scenario in which grid supply has been lost. It is assumed that when grid supply is restored, the software will switch to a ``normal operating'' mode. The normal operating mode may also be a sophisticated controller that seeks to, for instance, minimize the utility bill of the consumer by controlling the PV+battery system. There is a plethora of work in that direction; see~\cite{DiEvent:2012},~\cite{AnvariOptimal:2014},~\cite{BrahmanOptimal:2015},~\cite{MarzbandOptimal:2017} and~\cite{SanjariAnalytical:2017}. Therefore we do not consider that problem here. Works on controlling the PV+battery system to maximize resiliency performance in a post-disaster scenario, the focus of this paper, is extremely limited. To the best of our knowledge, only~\cite{PrinceResilience:2019} considers the problem of operation for resiliency. However, \cite{PrinceResilience:2019} ignores the mixed-integer nature of the optimization problem, and ignores the capability of a battery to vary charging rate which can be exploited during contingency situations like power outage.

Performance of the proposed control system is evaluated through simulations. The PV+battery system parameters are chosen according to existing design guidelines for standalone PV+battery systems as given in~\cite{MastersRenewable:2013}. The primary and secondary loads are chosen to be representative of what one might encounter in a typical home in Florida. For comparison, we also simulate a baseline controller that is representative of the existing commercial systems one can install today. It energizes both the primary and secondary loads' circuits if it estimates that there is enough energy available from the PV+battery system to service the combined demand. 

Simulations show that the proposed controller is able to service the primary load (refrigerator) throughout the simulation period (7 days after hurricane Irma in 2017) while the baseline controller is unable to do so for several hours each day. In addition, it is able to service the secondary load a little more than the baseline. We measure \emph{primary resiliency performance} of a control system as the average daily duration that the system is able to meet demand from the primary load. A simulation based study indicates that to meet a specific primary resiliency performance, the cost of the PV+battery system needed by the baseline controller is \emph{twice} that of that needed by the proposed controller. The cost of energy resiliency can therefore be halved by the proposed control system. 

The rest of the paper is organized as follows: the system description and mathematical models of the system components are described in Section~\ref{section:SystemDescription}. The formulation of proposed controller, and of a baseline controller, are provided in Section~\ref{section:ControlAlgorithms}. The simulation setup, computation and simulation parameters discussed in Section~\ref{section:SRD}. The results of the simulation study are presented and discussed in Section~\ref{section:ResultsDiscussions}. Finally, the main conclusions are provided in Section~\ref{section:Conclusions}.

\section{System Description and Models}\label{section:SystemDescription}
\begin{figure}[t]
	\centering
	\includegraphics[scale=0.3]{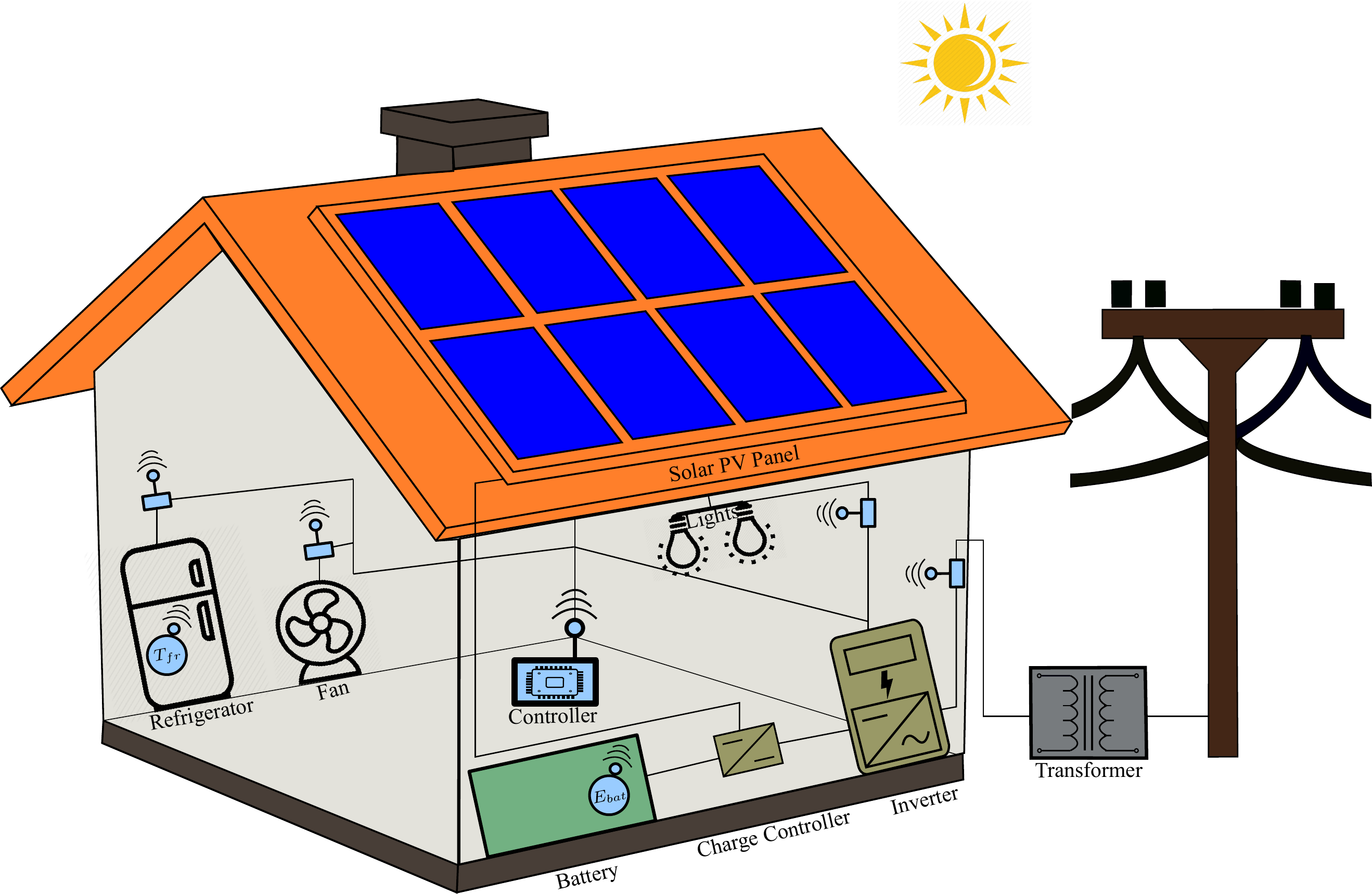}
	\caption{Hardware involved in the proposed control system.}
	\label{fig:PhysicalPlant_New}
\end{figure}

Figure~\ref{fig:PhysicalPlant_New} shows the schematic of a house with solar PV panels, a battery energy storage system, a  primary electric load (refrigerator), and secondary electric loads: lights and fans. 
The role of the proposed control system during a power outage is to control the following: (i) on/off state of the refrigerator, (ii) on/off state of the secondary load (aggregate of lights and fans),  (iii) charging/discharging state of the battery, and (iv) when charging the battery, the charging mode of  the battery. The battery has two charging modes: normal and fast. Fast is undesirable since it degrades battery life. The primary goal is to maintain the refrigerator temperature within the prescribed limits. A secondary goal is to service the secondary load during times that are pre-decided by the occupants. These goals are achieved by the controller using: (i) forecasted irradiance, (ii) estimated future house temperature, (iii) measured internal refrigerator temperature, and (iv) measured battery state of charge.

Mathematical models of each of these components are described in the subsections below. Time is discrete, with $k=0,1,2,\dots$ denoting the time index and $\Delta T_s$ denoting the interval (hours or minutes) between $k$ and $k+1$. In the sequel,  $E(k)$ ($Wh$) will denote the energy consumed/generated during the time interval between time indices $k$ and $k+1$, with the subscript specifying the source or consumer of the energy. The dependence on $k$ will be often omitted, e.g., we will say $x$ instead of $x(k)$. 
\subsection{Solar Photovoltaic Generation Model}\label{subsection:PVModel}
The PV output energy potential (this is the maximum energy the PV panels can produce at the current module temperature and available solar irradiance) is given by:
\begin{align} 
	E_{pv}(k) & =  N_{pv} \; P_{pv}^{rated} \; \left( \dfrac{G(k)} {G_{std}} \right)\times \;     \nonumber \\
	& \quad\left( 1 + \frac{\gamma}{100} \;\big(T_{m}(k)-T_{std}\big)  \right) \; \Delta T_{s}, \label{eq:PV_Model} 
\end{align} 
where $E_{pv}$ ($Wh$) is the energy output possible from the PV panels, $N_{pv}$ is the number of PV panels, $P_{pv}^{rated}$ ($W$) is the rated power output of PV module, $\gamma$ ($\%/^{\circ}C$) is the temperature coefficient of power of the PV module, $T_{m}$ ($^{\circ}C$) is the module temperature, $G_{std}$ ($W/m^{2}$) and $T_{std}$ ($^{\circ}C$) are the solar irradiance and ambient air temperature at standard test condition respectively, and $G$ ($W/m^{2}$) is the current solar irradiance. Eq.~\eqref{eq:PV_Model} is a modified version of that used in~\cite{TanakaOptimal:2012}; we use $P_{pv}^{rated}$ instead of PV conversion efficiency and array area which is equivalent, and $T_{m}$ instead of ambient temperature as it is a more accurate way of computing effect of ambient temperature on PV power~\cite{MastersRenewable:2013}.

The module temperature can be estimated from the ambient air temperature ($T_{am}$ in $^{\circ}C$) and wind speed ($W_s$ in $m/s$) using Faiman's formula~\cite{FaimanAssessing:2008} as follows:
\begin{align} 
	T_{m}(k)=T_{am}(k)+\dfrac{G(k)}{U_{0}+U_{1}+W_s(k)}, \label{eq:Faiman_Model} 
\end{align} 
where $U_{0}$ ($W/m^{2}K$) is the constant heat transfer component and $U_{1}$ ($W/m^{2}K$) is the convective heat transfer component.

\subsection{Battery Energy Storage System Model}\label{subsection:BatteryModel}
The battery storage system is simply modeled as a bucket of energy and its dynamics are as follows:
\begin{align} 
	E_{bat}(k+1) = E_{bat}(k)+\eta^{c}_{bat}E^{c}_{bat}(k)-\dfrac{E^{dc}_{bat}(k)}{\eta^{dc}_{bat}}, \label{eq:Battery_Model_1}
\end{align} 
where $E_{bat}$ ($Wh$) is the battery energy level, $E_{bat}^{c}~(Wh)$ is the energy absorbed (charging energy) by the battery to charge, and $E^{bat}_{dc}~(Wh)$ is the energy supplied (discharging energy) by the battery. Also, $\eta^{c}_{bat}$ and $\eta^{dc}_{bat}$ are the battery charging efficiency and the battery discharging efficiency, respectively. The battery energy level is bounded between the minimum ($\ubar E_{bat}$) and maximum ($\bar E_{bat}$) battery energy limits, i.e. $E_{bat}\in[\ubar E_{bat},\bar E_{bat}]$. 

The charging and discharging energies for the battery are constrained by the maximum charging and discharging energies as $E_{bat}^{c} \in [0,\bar E^{c}_{bat}]$ and $E_{bat}^{dc} \in [0,\bar E^{dc}_{bat}]$,  where $\bar E^{c}_{bat}~(Wh)$ and $\bar E^{dc}_{bat}~(Wh)$ are the maximum energies that the battery can absorb and supply during $\Delta T_s$, respectively. 

\subsection{Refrigerator Thermal Dynamic Model}\label{subsection:RefrigeratorThermalModel}
 We use the following discretized form of the continuous time refrigerator thermal dynamic model presented in~\cite{CostanzoGrey:2013}.
\begin{align} 
T_{fr}(k+1)=AT_{fr}(k) + Bu_{fr}(k) Q_{fr} + D T_{house}(k), 
\end{align}
where $T_{fr}~(^{\circ}C)$ is the internal refrigerator temperature, $u_{fr}$ is the refrigerator on-off control command, $Q_{fr}~(W)$ is the thermal power rejected by the refrigerator to the ambient when the compressor is on, and $T_{house}~(^{\circ}C)$ is the average internal house temperature. In addition, $Q_{fr}=COP \; P_{fr}^{rated}$, where $COP$ is the coefficient of performance, and $P_{fr}^{rated}$ is the rated power consumption of the refrigerator. $A$, $B$, and $D$ are the discrete time equivalents of the continuous time model given in \cite{CostanzoGrey:2013} and they are given by:
\begin{align*}
&A=e^{A_{c}\Delta T_{s}}, \quad B=\frac{1}{A_{c}}\left( e^{A_{c}\Delta T_{s}} - 1\right) B_c,\\
&D=\frac{1}{A_{c}}\left( e^{A_{c}\Delta T_{s}} - 1\right) D_c, 
\end{align*} 
where $A_{c}$, $B_{c}$, and $D_{c}$ are the continuous time constants of the model given as follows:
\begin{align*}
&A_{c}=\dfrac{-1}{C_{fr}R_{fr}}, \quad B_{c}=\dfrac{-1}{C_{fr}}, \quad D_{c}=\dfrac{1}{C_{fr}R_{fr}}, 
\end{align*}
where $R_{fr}~(^{\circ}C/W)$ and  $C_{fr}~(J/^{\circ}C)$ are the thermal resistance and thermal capacitance of the refrigerator respectively.

\subsection{Energy Consumption Models}\label{subsection:EnergyConsumptionModel}
The electrical energy consumed by the primary (refrigerator) and secondary (lights and fans) loads is simply the integral of their rated powers times the number of individual load units given as: $E_{fr}(k) = \left( P_{fr}^{rated}  \right) \; \Delta T_{s}$, $	E_{l}(k) = \left(  N_{l} \; P_{l}^{rated}  \right) \; \Delta T_{s}$, and $E_{f}(k) = \left(  N_{f} \; P_{f}^{rated}  \right) \; \Delta T_{s}$. Where $E_{fr}~(Wh)$, $E_{l}~(Wh)$ and $E_{f}~(Wh)$ are the electrical energies consumed by the refrigerator, lights and fans, respectively; $N_{l}$ and $N_{f}$ are the number of lights and fans respectively; $P_{fr}^{rated}~(W)$, $P_{l}^{rated}~(W)$ and $P_{f}^{rated}~(W)$ are the rated powers of the refrigerator, light and fan respectively. The energy consumed by the secondary loads, $E_{s}~(Wh)$ is given as follows:
\begin{align}\label{eq:Total_NC_Load}
E_{s}(k) = E_{l}(k) + E_{f}(k).
\end{align}   

\section{Control Algorithms and Plant}\label{section:ControlAlgorithms}
\begin{figure*}[t]
	\centering
	\includegraphics[scale=0.65]{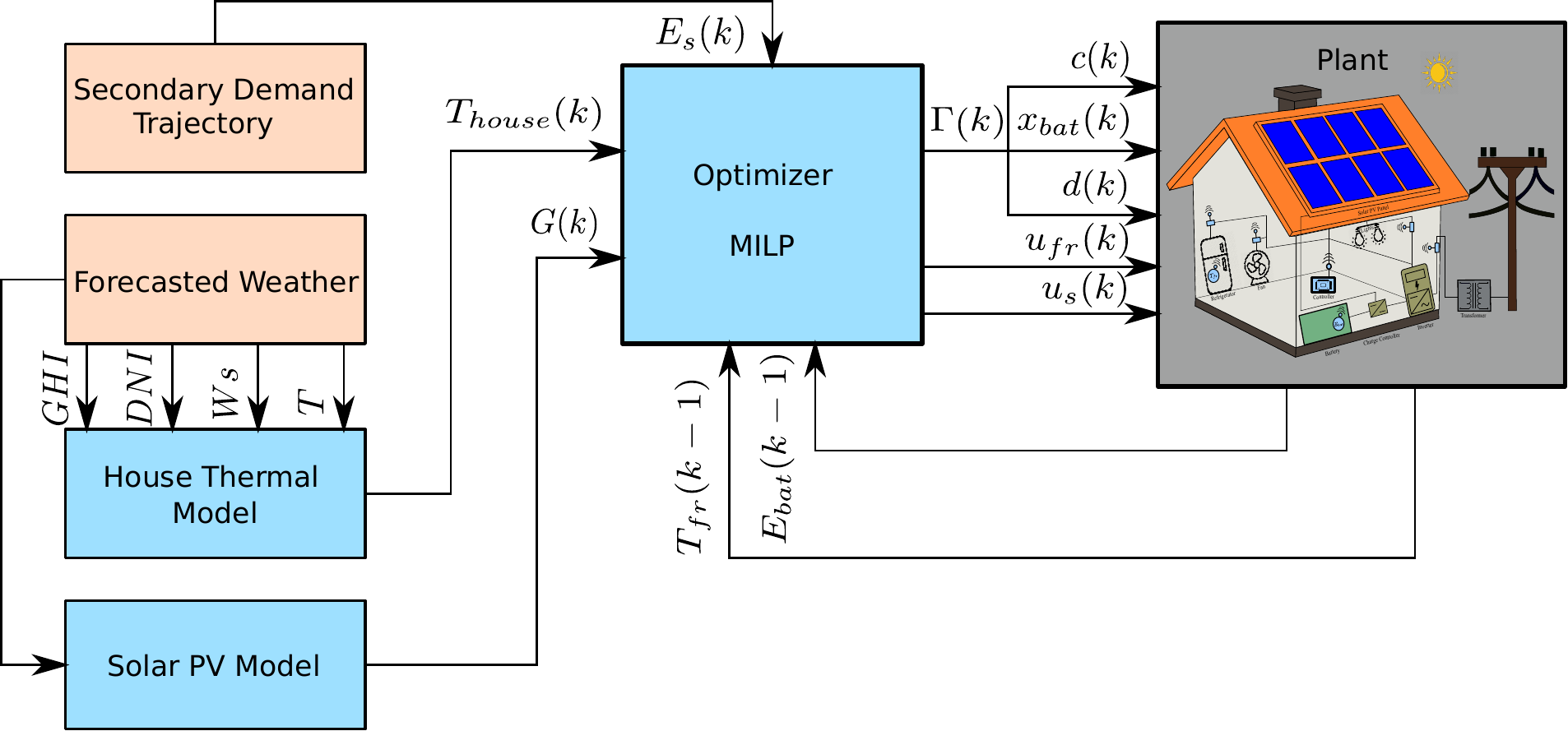}
	\caption{Schematic of closed loop operation of the proposed control system.}
	\label{fig:MILP_Schematic}
\end{figure*}
\subsection{Model Predictive Control (MPC)}\label{subsection:MPC}
The control decisions are computed at discrete time steps $k = 1,2 \dots N$  with $\Delta T_{s}$ as the sampling period, and $N$ is the total number of time steps in the planning/prediction horizon. The decision variables for the optimization problem elemental to the MPC controller are as follows: the states of the process $x(k) = [E_{bat}(k),T_{fr}(k)]^{T}$; the control commands $u(k) = [\Gamma(k),u_{fr}(k),u_{s}(k)]^{T}$, where $\Gamma(k)$ and $u_{s}(k)$ are the fraction of the normal battery charging energy and the secondary load on-off control command respectively; the internal variables $v(k) = [g(k),\zeta_{fr}(k)]^{T}$, where $g(k)$ is the energy produced by the PV panels between $k$ and $k+1$ time steps, and $\zeta^{fr}(k)$ is a slack variable for refrigerator temperature to ensure feasibility. The exogenous inputs whose predictions are assumed to be known for the $N$ time steps $w(k) = [G(k),T_{house}(k),E_{s}(k)]^{T}$, where $G(k)$ and $E_{s}(k)$ are the available energy from the PV panels computed using \eqref{eq:PV_Model} and the total secondary load computed from \eqref{eq:Total_NC_Load} respectively, and $T_{house}(k)$ is the internal house temperature. Hence, the complete decision vector for the optimization problem is given as $[X,U,V]^{T}$, where $X:=[x(k+1), \dots ,x(k+N)]^{T}$, $U:=[u(k), \dots , u(k+N-1)]^{T}$ and $V:=[v(k), \dots , v(k+N-1)]^{T}$.

A constrained optimization problem is solved, to generate control commands for N time steps, which tries to keep the refrigerator temperature in the prescribed temperature range, maximize the battery state of charge, minimize the health degradation of the battery, and maximize the operation of the secondary loads. This objective is achieved subject to constraints on the refrigerator and battery model dynamics, energy balance equation, battery state of charge constraints, and the constraints on charging and discharging rate of the battery. The optimization problem at any time index $j$ is given mathematically as follows:
\begin{equation}
\begin{aligned} 
	\min_{X,U,V} & \sum\limits^{j+N-1}_{k=j} \bigg[\lambda_{1}(N-k)\zeta_{fr}(k) - \lambda_{2}E_{bat}(k) + \\ 
	&\quad \quad \quad \quad \lambda_{3}\Gamma(k) - \lambda_{4}(N-k)u_{s}(k) \bigg], \label{eq:CostFunction}
\end{aligned}
\end{equation}
subject to the following constraints:
\begin{align}
	&T_{fr}(k+1)=AT_{fr}(k) + Bu_{fr}(k) Q_{fr} + D T_{house}(k), \label{eq:Fridge_eqcon_1} \\
	&E_{bat}(k+1) = E_{bat}(k) + \Gamma(k)  \eta_{bat}^{c,dc,con} \bar E_{bat}^{c}, \label{eq:Battery_eqcon_1}\\
	&u_{fr}(k) E_{fr} + \Gamma(k) \bar E_{bat}^{c} + u_{s}(k)E_{s}(k)=  g(k), \label{eq:EnergyBalance_eqcon_1} \\
	&\ubar T_{fr}\leq T_{fr}(k)\leq \bar T_{fr} + \zeta_{fr}(k), \label{eq:Fridge_ineq_1} \\
	&\zeta_{fr}(k)\geq 0, \label{eq:Slack_ineq_1} \\
	&\ubar E_{bat}\leq E_{bat}(k)\leq \bar E_{bat}, \label{eq:Battery_ineq_1} \\
	&\ubar u_{s} \leq u_{s}(k) \leq \bar u_{s}(k), \label{eq:NCControl_ineq_1} \\
	& \ubar\Gamma \leq \Gamma(k) \leq \bar\Gamma, \label{eq:Battery_ineq_2} \\
    &0 \leq g(k) \leq G(k). \label{eq:PV_ineq_1}
\end{align}

The cost function in (\ref{eq:CostFunction}) consists of four terms which helps in achieving the objectives of the optimization problem. The first term, $\lambda_{1}(N-k)\zeta_{fr}(k)$, penalizes the refrigerator temperature slack variable which in turn tries to keep the refrigerator internal temperature within prescribed bounds. The time varying weighing factor $N-k$ puts a higher penalty on the slack at earlier time and less weight on later times, which in turn translates to having a smaller slack during the initial time steps of the planning horizon. The second term, $- \lambda_{2}E_{bat}(k)$, penalizes a low state of charge which helps in extending the life-time of the system. The third term, $\lambda_{3}\Gamma(k)$, puts a higher penalty on larger battery charging rate. $\Gamma$ models the fraction of the charging/discharging energy of the battery which are continuous and variable; moreover, it depends on the amount of energy available from the PV panels during charging, and the amount of the load demand to be supplied during discharging. This penalty is added to encourage normal charging instead of faster charging, since faster charging reduces battery health. The fourth term, $- \lambda_{4}(N-k)u_{s}(k)$, along with the inequality constraint \eqref{eq:NCControl_ineq_1} maximizes the operation of the secondary loads when desired. The reason for the time varying weight in this term is similar to that in the first term. The parameters $\lambda_{1}, \lambda_{2}, \lambda_{3}$, and $\lambda_{4}$ are designer specified wights and have to be selected carefully to derive the desired response from the optimizer. 

The equality constraint (\ref{eq:Fridge_eqcon_1}) is due to the thermal dynamics of the refrigerator. The equality constraint \eqref{eq:Battery_eqcon_1} is due to the battery energy dynamics, where $\bar E_{bat}^{c}$ is the maximum battery charging and discharging energy (which are assumed to be equal for modeling simplicity) in the normal mode, and $\eta_{bat}^{c,dc,con}$ is the charging-discharging efficiency of battery used in the controller. This battery dynamics differ from plant battery dynamics given in eq.~\eqref{eq:Battery_Model_1} as it models the battery charging and discharging energies with a single continuous variable ($\Gamma$). The equality constraint \eqref{eq:EnergyBalance_eqcon_1} is the energy balance equation, where $E_{fr}$ is the energy consumed by the refrigerator between $k$ and $k+1$ time steps.

The inequality constraint \eqref{eq:Fridge_ineq_1} is to maintain the refrigerator temperature within the lower ($\ubar T_{fr}$) and upper ($\bar T_{fr}$) temperature limits. The inequality constraint \eqref{eq:Slack_ineq_1} is present to not allow the refrigerator temperature slack to become negative. The inequality constraint \eqref{eq:Battery_ineq_1} bounds the battery energy between the minimum ($\ubar E_{bat}$) and maximum ($\bar E_{bat}$) battery energy limits. The inequality constraint \eqref{eq:NCControl_ineq_1} is present to force the secondary load control command to be zero when secondary loads are not desired to be turned on by the occupants, where $\ubar u_{s}$ and $\bar u_{s}$ are the lower and upper bound on $u_{s}$ respectively, and are defined as follows:
\begin{align}
\bar u_{s}(k)= & 
\begin{cases}
1 \; , & \text{if } E_{s}(k) > 0 \\
0 \; , & \text{if } E_{s}(k) = 0 \\
\end{cases}\\
\ubar u_{s}(k)= & 0, \forall\;\; k = 1,2,\dots,N.
\end{align}
The inequality constraint (\ref{eq:Battery_ineq_2}) bounds the fraction of battery charging/discharging energy ($\Gamma$) between a minimum ($\ubar\Gamma=-1$) and maximum ($\bar \Gamma=2$) value. For negative values the battery discharges; whereas for positive values till 1 it charges in normal mode with a battery charging energy of $\bar E_{bat}^{c}$ as the maximum, for values above 1 it charges in fast mode with twice the normal battery charging energy of $2 \times \bar E_{bat}^{c}$ as the maximum. The inequality constraint \eqref{eq:PV_ineq_1} bounds the energy produced by the PV panels such that it cannot be negative and is always less than or equal to the available PV energy ($G$).

The control variables $u_{fr}$ and $u_{s}$ are modeled as binary integer variables, taking values in $\{1,0\}$ to turn the loads on and off respectively. This binary nature of the control commands makes this problem a Mixed Integer Linear Program (MILP).

\subsection{Overall Plant Model}\label{subsection:PlantModel}
The plant used for closed loop simulations consists of the dynamic models presented in Section~\ref{section:SystemDescription}, and the interactions between the PV panels, the battery, and the loads (both primary and secondary). These interactions are represented mathematically using the following equations:
\begin{align}
	&E_{pv}(k) = E_{pv}^{u}(k)+E_{pv}^{un}(k), \label{eq:PlantModel_1} \\
	&E_{hl}(k) = \dfrac{u_{fr}(k)E_{fr}(k)+u_{s}(k)E_{s}(k)}{\eta_{inv}}, \label{eq:PlantModel_2} \\
	&E_{pv}^{u}(k) = E_{hl}(k)+E^{c}_{bat}(k), \label{eq:PlantModel_3} \\
	&E^{c}_{bat}(k) = c(k) \; min\big\{E_{pv}(k)-E_{hl}(k), \nonumber \\ 
	& \quad \quad \quad \quad \quad \bar E_{bat}-E_{bat}(k),  x_{bat}(k) \; \bar E^{c}_{bat} \big\}, \label{eq:PlantModel_5} \\
	&E^{dc}_{bat}(k) = d(k) \; min\big\{E_{hl}(k)-E_{pv}(k), \nonumber \\  
	& \quad \quad \quad \quad \quad E_{bat}(k)-\ubar E_{bat}, \bar E^{dc}_{bat}\big\}. \label{eq:PlantModel_6}
\end{align}
Eq.~\eqref{eq:PlantModel_1} shows that the maximum available energy that can be produced by the PV panels ($E_{pv}$) is equal to the PV energy used ($E_{pv}^{u}$) and PV energy unused ($E_{pv}^{un}$) between the time steps $k$ and $k+1$. Eq.~\eqref{eq:PlantModel_2} shows that the total house load ($E_{hl}$) is composed of the energy used by the refrigerator ($E_{fr}$) and the secondary loads ($E_{s}$) between the time steps $k$ and $k+1$, where $\eta_{inv}$ is the inverter efficiency. Eq.~\eqref{eq:PlantModel_3} shows that the PV energy used supplies the total house load and battery charging energy ($E^{c}_{bat}$) between the time steps $k$ and $k+1$. Eq.~\eqref{eq:PlantModel_5} gives the battery charging energy such that it never charges beyond the maximum battery energy limit ($\bar E_{bat}$), where $x_{bat}$ is either 1 or 2 (1 - normal charging and 2 - fast charging), $\bar E^{c}_{bat}$ is the maximum battery charging energy and $c$ is the battery charging control command ($c=1$ for charging, $c=0$ for not charging). Eq.~\eqref{eq:PlantModel_6} gives the battery discharging energy such that it never discharges below the minimum battery energy limit ($\ubar E_{bat}$), where $\bar E^{dc}_{bat}$ is the maximum battery discharging energy, and $d$ is the battery discharging control command ($d=1$ for discharging, $d=0$ for not discharging). 

\subsubsection{Implementation of the Control Commands Computed by the MPC Controller in the Plant}\label{subsubsection:DwellTime}
Figure~\ref{fig:MILP_Schematic} shows the control architecture for the proposed MPC controller. The control commands $ u_{fr}$ and $u_{s}$ are directly applied to the plant, turning the refrigerator and the secondary loads on and off depending on whether $ u_{fr}$ and $u_{s}$ are 1 and 0 respectively. However, $\Gamma$ is converted into appropriate discrete decisions, $c$, $d$ and $x_{bat}$, which are then applied to the charge controller in the following manner:
\begin{align}
	c(k)= & 
	\begin{cases}
		1 \; , & \text{if } \Gamma(k) > 0 \\
		0 \; , & \text{if } \Gamma(k) \leq 0 \\
	\end{cases}\\
	d(k)= &
	\begin{cases}
		1 \; , & \text{if } \Gamma(k) < 0 \\
		0 \; , & \text{if } \Gamma(k) \geq 0 \\
	\end{cases}\\
	x_{bat}(k)= &
	\begin{cases}
		1 \; , & \text{if } 0 < \Gamma(k) \leq 1 \\
		2 \; , & \text{if } 1 < \Gamma(k) \leq 2 \\
		0 \; , & \text{otherwise } .  \\
	\end{cases}
\end{align}

\subsection{Baseline Controller}\label{subsection:Baseline}
 The baseline controller consists of two independent controllers: refrigerator dead-band controller and the charge controller.  The naive dead-band controller controls the on-off ($u_{fr}$) of the refrigerator as follows: 
\begin{align}
	u_{fr}(k+1)= 
	\begin{cases}
		1,& \text{if }  T_{fr}(k+1) \geq \bar T_{fr}\\
		0,& \text{if }  T_{fr}(k+1) \leq \ubar T_{fr}\\
		u_{fr}(k),& \text{otherwise, }
	\end{cases}
\end{align}
where $\bar T_{fr}$ and $\ubar T_{fr}$ are the maximum and minimum refrigerator temperature limits respectively. The charging ($c$) and discharging ($d$) of the battery is controlled by the charge controller as: 
\begin{align}
c(k)= 
\begin{cases}
1,& \text{if } E_{pv}(k) >  E_{hl}(k) \\
0,& \text{otherwise }  \\
\end{cases}\\
d(k)= 
\begin{cases}
1,& \text{if } E_{pv}(k) <  E_{hl}(k) \\
0,& \text{otherwise. } \\
\end{cases}
\end{align}
However, the amount of charging energy depends on the surplus energy production from the PV panels after the house loads have been serviced and the battery energy level as given in \eqref{eq:PlantModel_5}; and the amount of discharge energy depends on the house load energy not served by the PV panels and the battery energy level as given in \eqref{eq:PlantModel_6}. The baseline controller services load demand till the system has adequate energy. It has only the normal charging mode.


\section{Simulation Study Setup}\label{section:SRD}
The period selected for simulation is the time hurricane Irma passed over Gainesville, FL, USA, starting from its landfall on Sept. 11, 2017, to Sept. 17, 2017. Weather data is obtained from National Solar Radiation Database (\url{nsrdb.nrel.gov}). The simulations are run for 7 days starting at 00:00~hours (midnight) at day 1 (September 11, 2017)  with a planning horizon of 24 hours and a time step of 10 minutes ($\Delta T_{s}=10 \text{ mins}$, $N=144$) with battery initial state at $\bar E_{bat}$ (i.e., $E_{bat}(0)=\bar E_{bat}$) and the refrigerator initial temperature at 2$^{\circ}C$ (i.e., $T_{fr}(0)=2^{\circ}C$). The internal house temperature, $T_{house}(k)$, for the planning horizon is computed using the linear model given by~\cite{CuiHybrid:2019}, which models a typical, detached, two-story house in the USA. 

\subsection{PV Battery System Sizing}\label{subsection:SystemSizing}
The PV battery system sizing is done using the method given in~\cite{MastersRenewable:2013}. This method requires specification of PV panel type and battery type, inverter efficiency, average daily energy demand, average daily solar insolation, and the desired number of days of energy storage. The method then computes the number of PV panels required, number of battery units required, and their respective configuration (number of PV panels/battery units in series and parallel), and the system DC voltage required to supply the load for the selected number of storage days in presence of the given average daily solar insolation. We selected the Canadian Solar CS6K-285 polycrystalline panel ($\$100/panel$), and Trojan SPRE 12 225 (lead acid type) solar battery unit ($\$400/unit$). Lead acid battery is selected over Lithium-Ion (Li-ion) battery  despite the latter having performance advantages over the former in order to reduce cost, since Li-ion batteries are four times more expensive than lead acid batteries per $kWh$~\cite{DioufThe:2019}. The desired number of days of storage was selected to be 1, even though we wish the system to be able to meet the primary and secondary loads for several days after an outage. This was again done in the interest of cost, since otherwise a large number of PV panels and a large battery will be required.

\begin{figure}[htpb]
	\centering
	\includegraphics[scale=0.5]{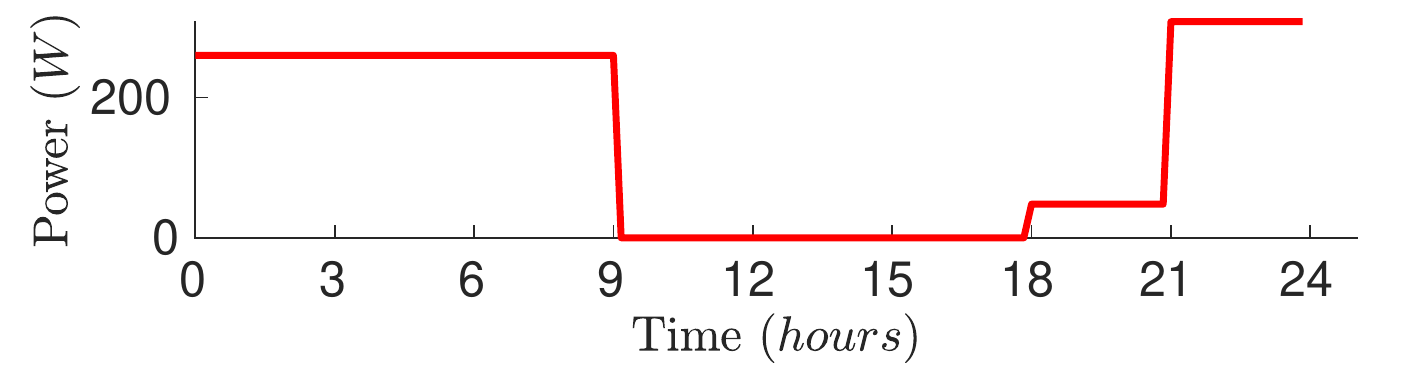}
	\caption{Secondary load demand (daily trajectory).}
	\label{fig:NCTrajectory}
\end{figure}
The house described in \cite{CuiHybrid:2019} consists of four bedrooms, a living room, and a kitchen. Hence, during and post hurricane period when power from grid is not available, the minimum load which will provide habitable conditions was decided to be an LED light for each room, a fan for each bedroom, and one refrigerator in the kitchen. Fig.~\ref{fig:NCTrajectory} illustrates the secondary load trajectory for a given day which is composed of: LED lights being on from 18:00~hours to 00:00~hours and fans running from 21:00~hours to 09:00~hours. 
The size of the system obtained from this method is as follows: 3 PV panels connected in parallel, 2 units of battery connected in series and the system DC voltage ($V_{d}$) is 24$V$.

\subsection{Computation}\label{subsection:Computation}
The plant is simulated in MATLAB. The optimization problem is solved using GUROBI~\cite{gurobi:2019}, a mixed integer linear programming solver, on a Desktop Linux computer with 8GB RAM and a 3.60 GHz $\times$ 8 CPU. On an average it takes 60 seconds for GUROBI solver to solve the MILP for one planning horizon. Moreover, the default setting for the MIPGap (optimality gap) option which is 0.01\% is too high for this problem, and causes the solver to stall in search for an optimal solution to meet this tolerance. A more practical value of 1\% is used for MIPGap. Other changes to the default settings are as follows: MIPFocus = 3 (focus on bound), Method = 4 (deterministic concurrent), and Cuts = 3 (aggressive cut generation). Even with all the above changes there can be some iterations of the MPC where the solver struggles to meet the relaxed MIPGap and stalls. For this situation, the option TimeLimit is used to terminate the solver after 5 minutes of solution search. In this case, the best solution found by the solver at the end of the 5 min period is used for control. The solver stalled for 7.5\% of the times.

\subsection{Simulation Parameters}\label{subsection:SimulationParameters}
The parameters for the plant components; PV panels: $N_{pv}=3$ and $P_{pv}^{rated}=285$ $W$, $\gamma=-0.39$ $\%/^{\circ}C$, $G_{std}=1000$ $W/m^{2}$, $T_{std}=25$ $^{\circ}C$, $U_{0}=25$ $W/m^{2}$$\degree C$ and $U_{1}=6.84$ $W/m^{2}$$\degree C$; Battery: $\ubar E_{bat}=1080$ $Wh$, $\bar E_{bat}=5400$ $Wh$, $\bar E_{bat}^{c}=810$ $Wh$, $\bar E_{bat}^{dc}=844.5$ $Wh$, $\eta^{c}_{bat}=0.9$ and $\eta_{bat}^{dc}=0.9$ and Loads: Refrigerator - $P_{fr}^{rated}=250$ $W$, $\ubar T_{fr}=0$ $^{\circ}C$, $\bar T_{fr}=4$ $^{\circ}C$, Lights - $N_{l}=6$, $P_{l}^{rated}=8$ $W$ Fans - $N_{f}=4$, $P_{f}^{rated}=65$ $W$. The system voltage is $V_{d}=24$ $V$ and the invert efficiency is $\eta_{inv}=0.9$.

The parameters for the refrigerator thermal model are $C_{fr}=8.9374\times10^{3}$ $J/^{\circ}C$, $R_{fr}=1.4749$ $^{\circ}C/W$ and $COP=0.2324$. 

The parameters for the optimization problem are $\lambda_{1}=1$, $\lambda_{2}=1$, $\lambda_{3}=1$, $\lambda_{4}=10$, $\eta_{bat}^{c,dc,con}=1$, $\ubar\Gamma=-1$ and $\bar\Gamma=2$.

\section{Results and Discussion}\label{section:ResultsDiscussions}
\begin{figure*}[!t]
	\centering
	\begin{subfigure}[t]{0.48\textwidth}
		\centering
		\includegraphics[scale=0.45]{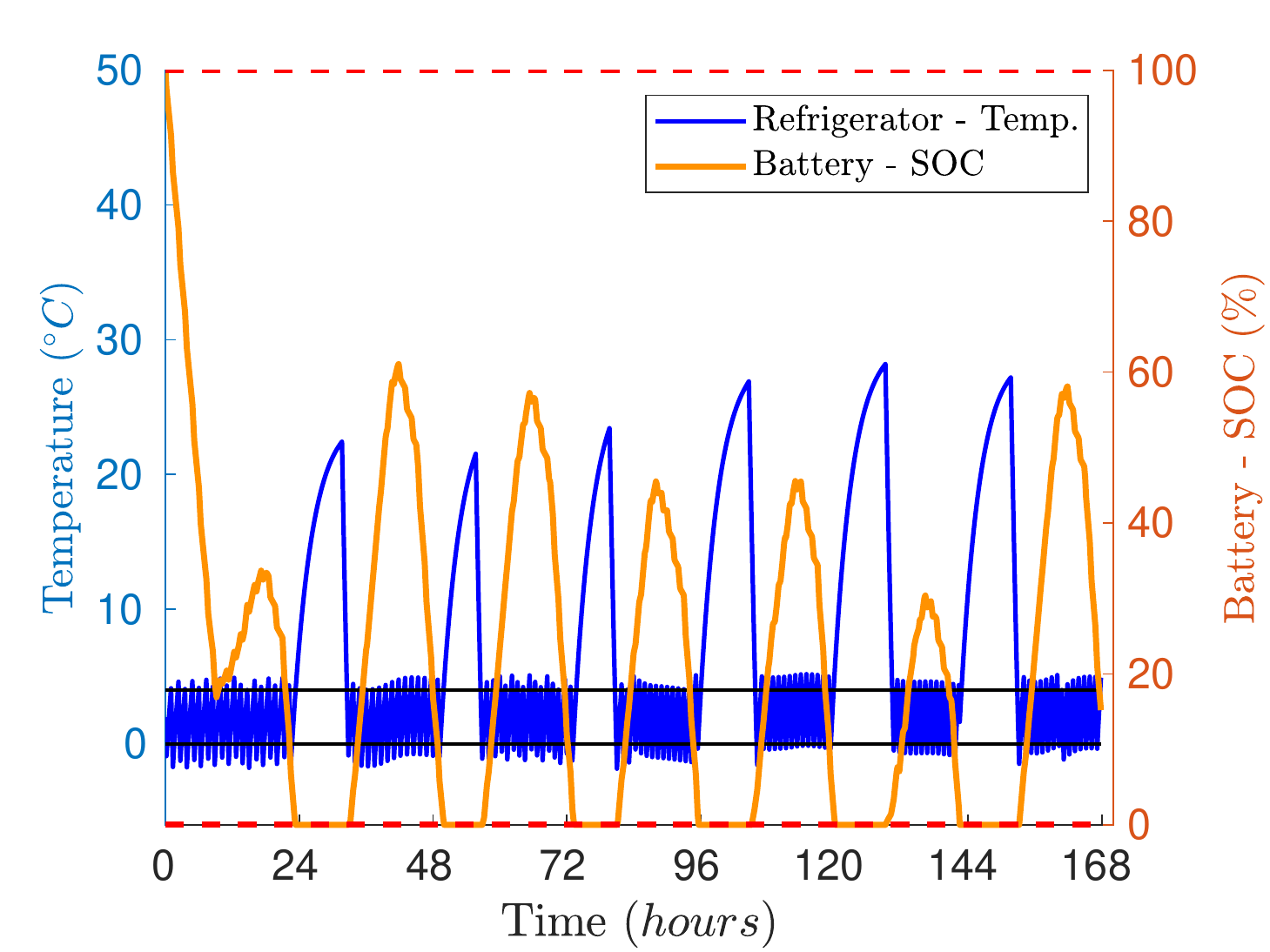}
		\caption{Baseline controller: Refrigerator temperature and Battery-SoC.}
		\label{fig:Result_1}
	\end{subfigure}
	\begin{subfigure}[t]{0.48\textwidth}
		\centering
		\includegraphics[scale=0.45]{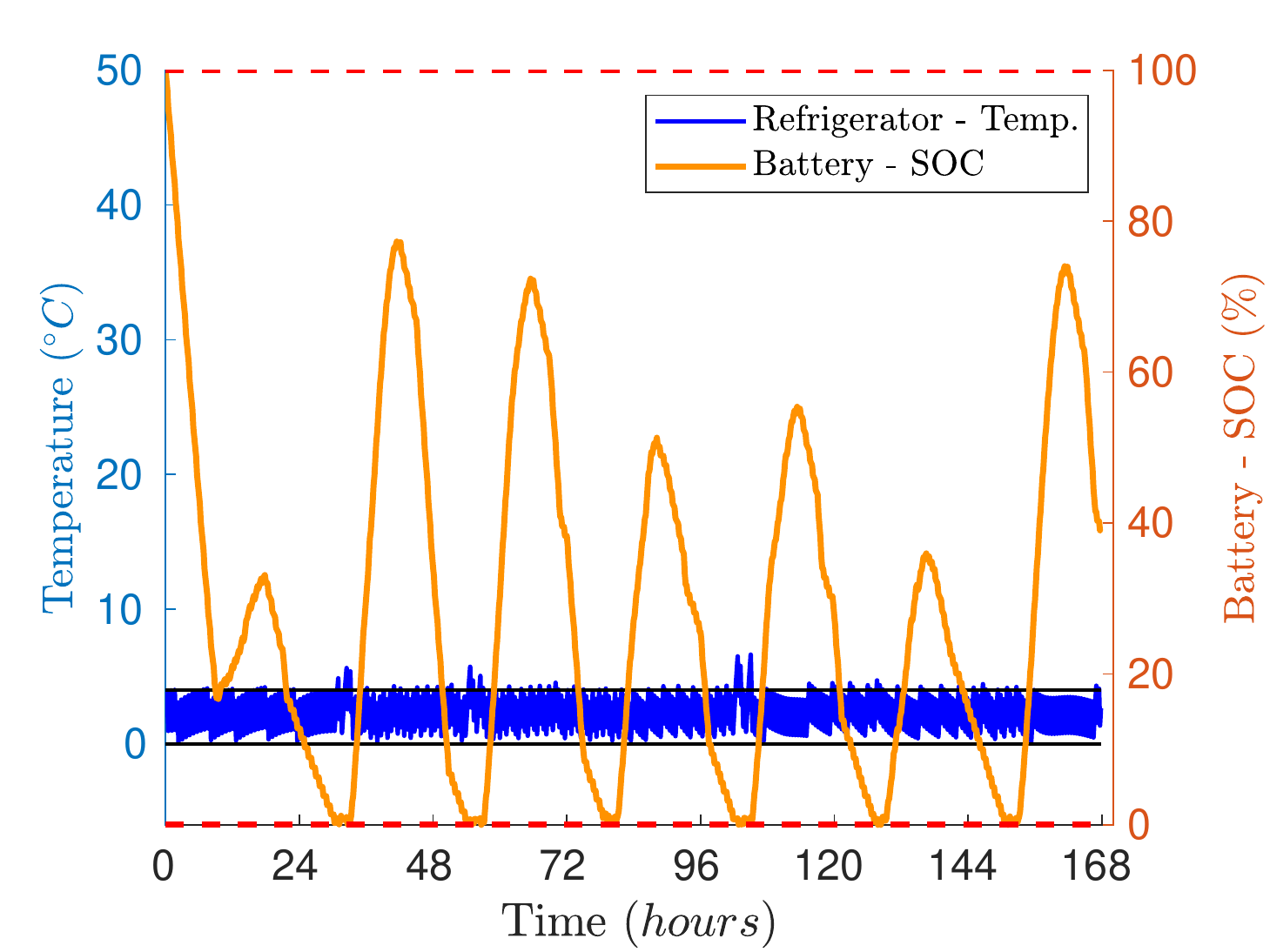}
		\caption{Proposed controller: Refrigerator temperature and Battery-SoC.}
		\label{fig:Result_2}
	\end{subfigure}
	
	\begin{subfigure}[t]{0.48\textwidth}
		\centering
		\includegraphics[scale=0.45]{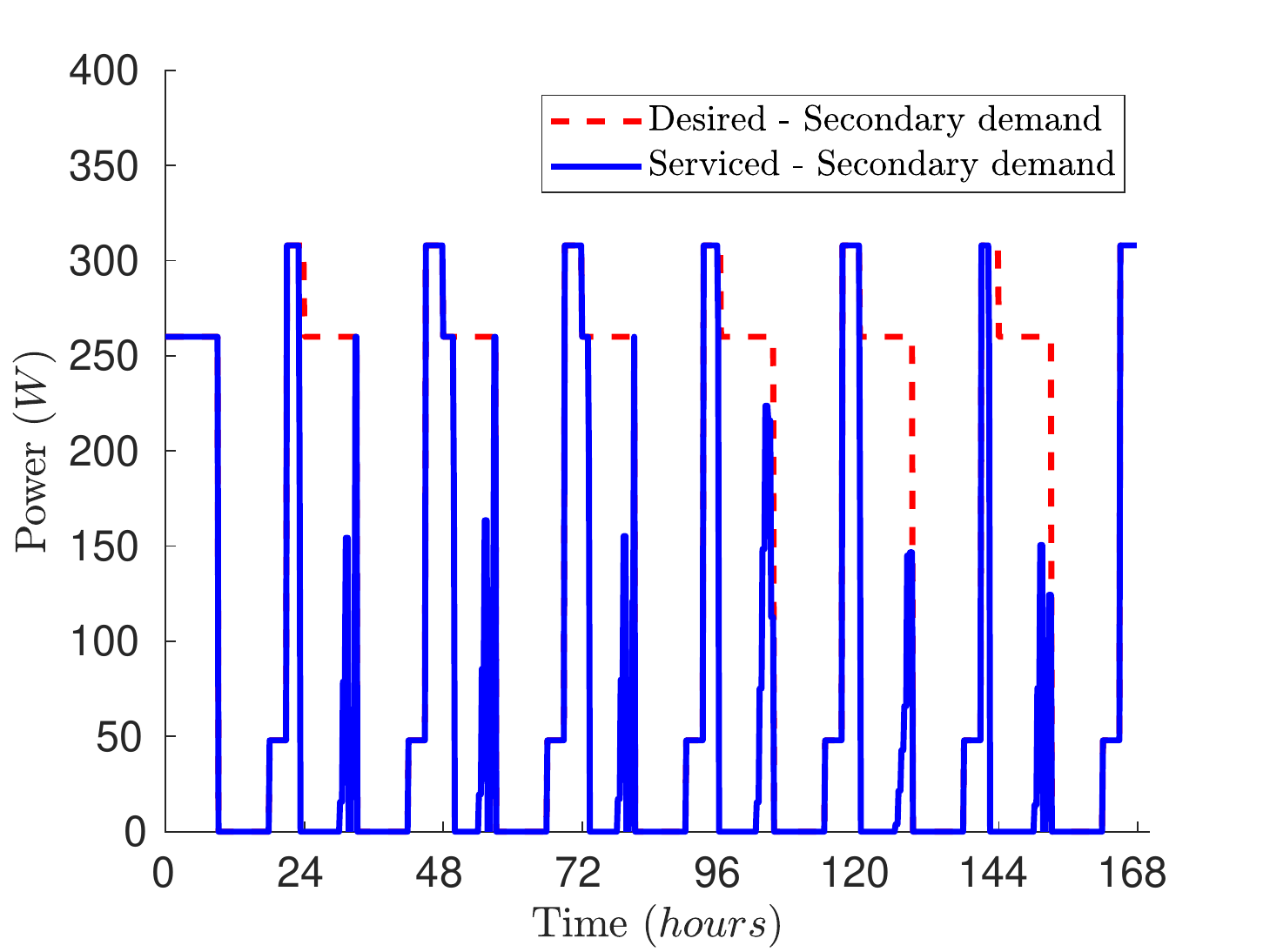}
		\caption{Baseline controller: secondary demand meeting performance.}
		\label{fig:Result_3}
	\end{subfigure} 
	\begin{subfigure}[t]{0.48\textwidth}
		\centering
		\includegraphics[scale=0.45]{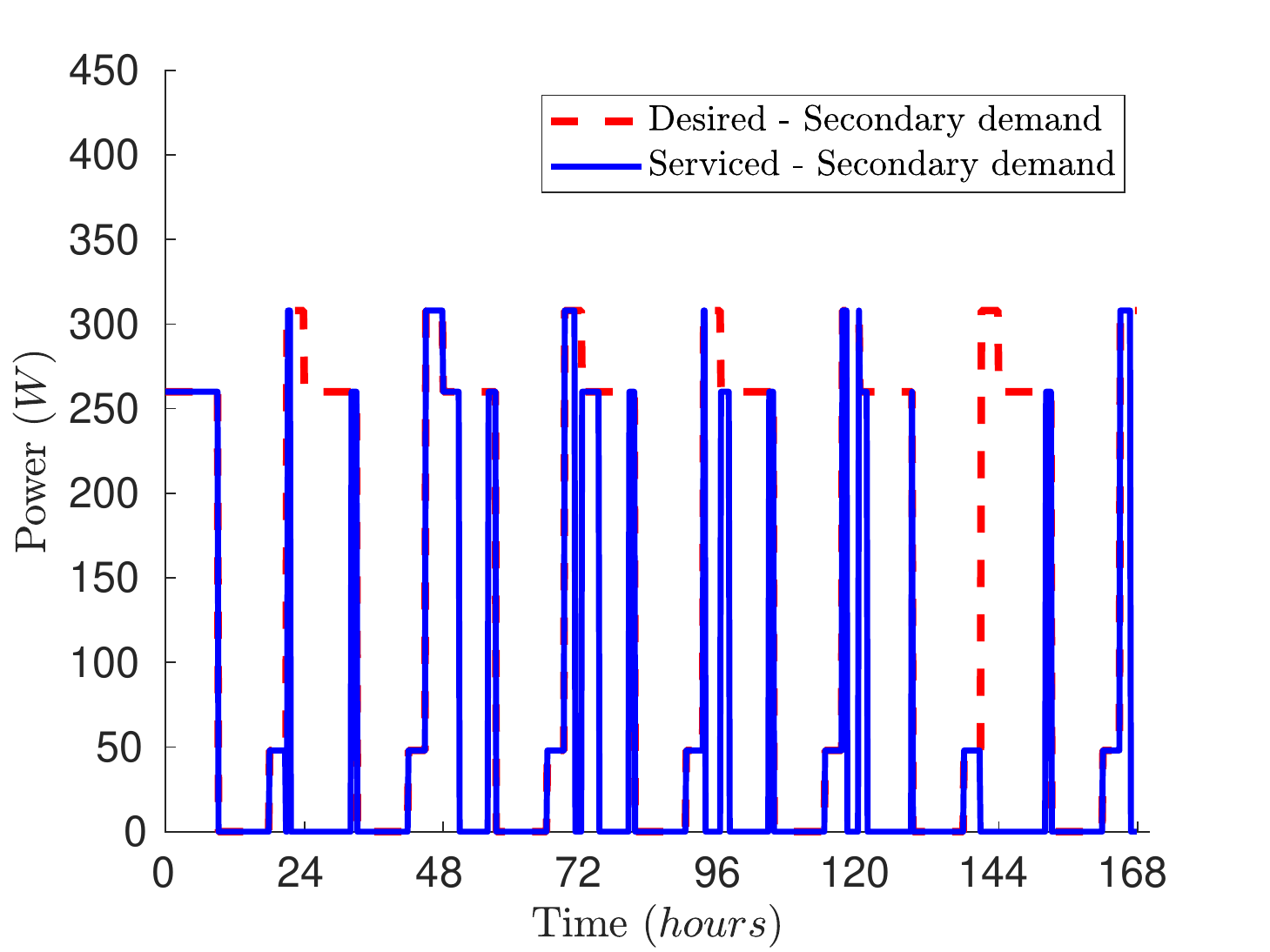}
		\caption{Proposed controller: secondary demand meeting performance.}
		\label{fig:Result_4}
	\end{subfigure} 
	\caption{ Comparison of Baseline and Proposed controllers' performances for the week after hurricane Irma in Gainesville, FL.}
	\label{fig:Results}
\end{figure*}
\begin{figure}[htpb]
	\centering
	\includegraphics[scale=0.5]{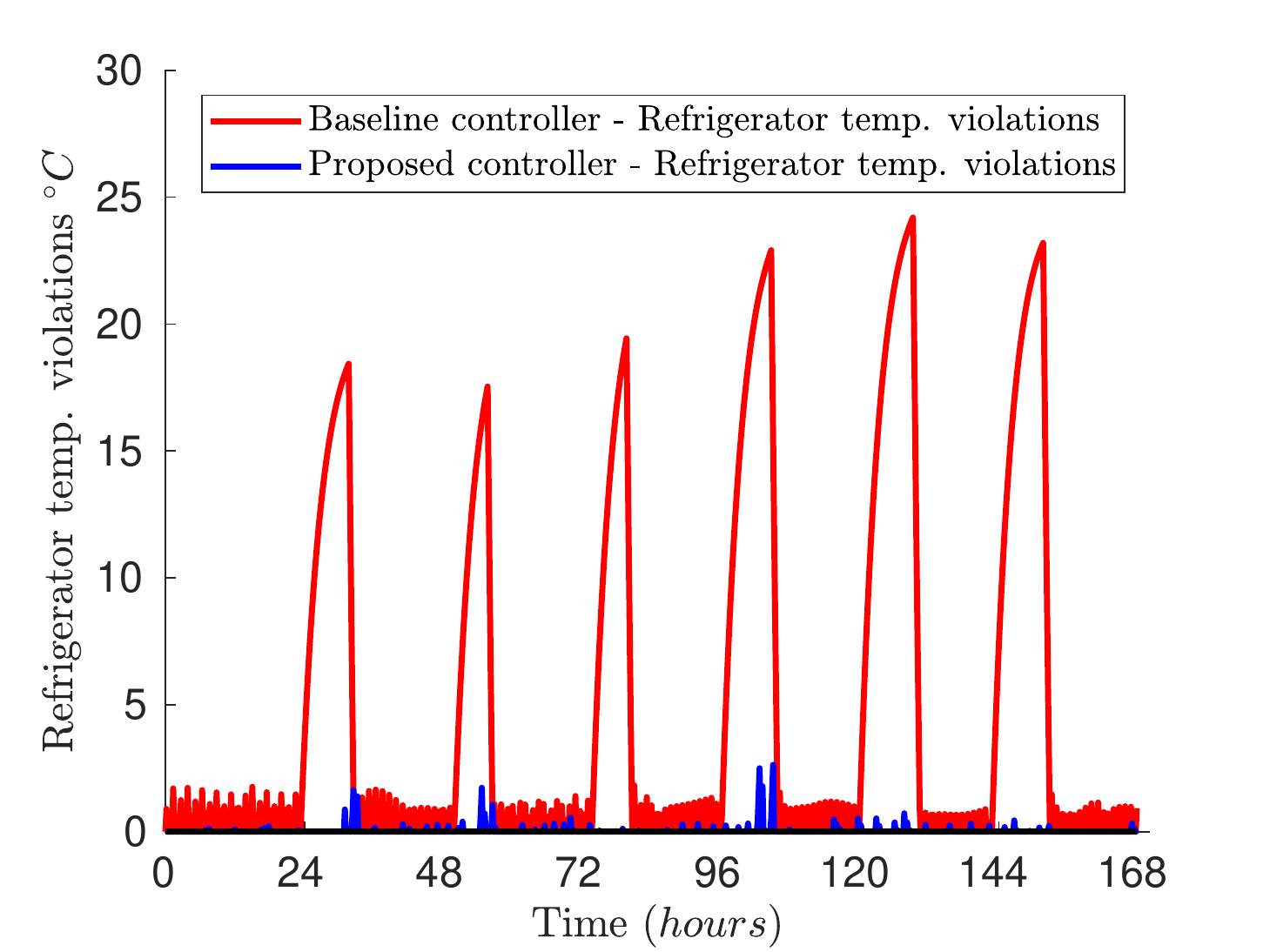}
	\caption{Comparison of baseline and proposed controllers' refrigerator temperature violations from the dead-band.}
	\label{fig:Result_BaslineMILP_TempViolation}
\end{figure}

Figure~\ref{fig:Results} shows the simulation results when using the baseline and the proposed controllers.
The proposed controller keeps the refrigerator temperature within the prescribed limits for the entire 7 days with minor excursions; see Figure~\ref{fig:Result_2}. In contrast, the baseline controller fails to do so for elongated periods; see Figure~\ref{fig:Result_1}. Figure~\ref{fig:Result_BaslineMILP_TempViolation} shows the amount of  refrigerator temperature violation (distance to the allowed band) incurred by each of the controllers. 
The average daily refrigerator temperature violation is  7.13~hours/day for the baseline controller, but only 0.04~hours/day for the proposed controller; see Table~\ref{tab:Results}.  Note that the Centers for Disease Control and Prevention state that perishable foods (including meat, poultry, fish, eggs and leftovers) in the refrigerator should be thrown away if the power has been off for 4 hours or more~\cite{CdcKeep:2019}. Thus, while the proposed controller will be able to keep perishable foods fresh for the entire seven days of the outage, with the baseline controller, the stored food will get spoiled after the very first day without grid power.

Figures~\ref{fig:Result_3} and \ref{fig:Result_4} show the trajectories of the secondary loads serviced by the baseline and proposed controllers respectively. It can be seen that none of the controllers are able to meet the secondary loads for the desired duration. However, the proposed controller has a slightly better performance (8\%) than the baseline controller even in case of servicing the secondary loads; see Table~\ref{tab:Results}.

Hence, the proposed controller demonstrates superior performance in servicing both primary and secondary loads as compared to the baseline controller. The superior performance of the proposed controller is attributed to (i) its taking into account forecasts of disturbances (solar energy available, temperature of the house, and desired trajectory for the secondary loads) in making decisions and (ii) making the trade off between various conflicting requirements by solving an optimization problem. On the contrary, the baseline controller operates with the information consisting of just the present states ($T_{fr}(k)$ and $E_{bat}(k)$) of the system and its decision making is simple (rule based).
\begin{table}[b]
	\caption{Performance comparison of baseline and proposed controller.}
	\label{tab:Results}
	\begin{center}
		\begin{tabular}{|c||c||c|}
			\hline
			& Baseline & Proposed\\
			\hline
			Refrigerator temp. violation ($hours/Day$) & 7.1250 & 0.0416\\
			\hline
			Secondary loads not served (\% $time$) & 57 & 48.63\\
			\hline			
		\end{tabular}
	\end{center}
\end{table} 

\begin{table}[h]
	\caption{System size, description, and cost.}
	\label{tab:SystemSize}
	\begin{center}
		\begin{tabular}{|c||c||c|}
			\hline
			System size& Description & Cost\\
			\hline
			A & 3 PV panels + 2 Battery units  & \$1100\\
			\hline
			B & 4 PV panels + 2 Battery units  & \$1200\\
			\hline
			C &  3 PV panels + 4 Battery units  & \$1900\\
			\hline
			D & 4 PV panels + 4 Battery units  & \$2000\\
			\hline
			E & 5 PV panels + 4 Battery units  & \$2100\\
			\hline
			F & 6 PV panels + 4 Battery units  & \$2200\\
			\hline			
		\end{tabular}
	\end{center}
\end{table}

To observe the effect of system size on the performance of the two controllers, the baseline controller was simulated with six different system sizes. Table~\ref{tab:SystemSize} shows the different system sizes which were used for the simulations in ascending order of size/cost. For each system the PV panels are connected in parallel, and the battery units are connected in a series-string of 2 units/string, any extra addition of battery units has to be in a series-string of 2 units/string connected in parallel to the previous series strings.

\begin{figure}[htpb]
	\centering
	\includegraphics[scale=0.6]{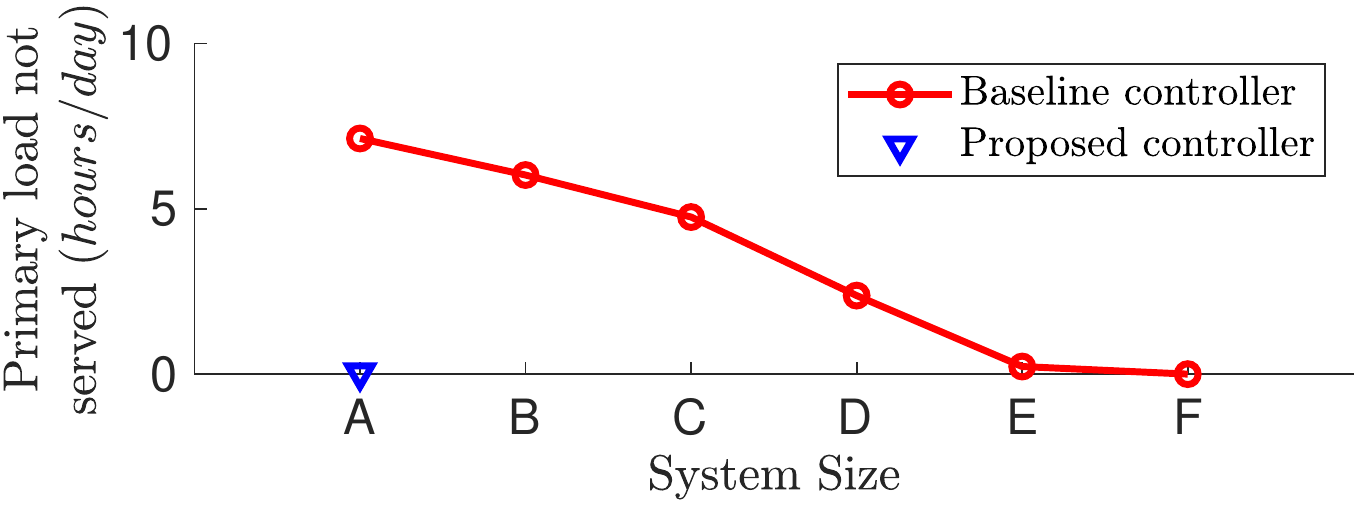}
	\caption{Comparison of baseline and proposed controllers' resiliency performance with increase in system size.} 
	\label{fig:Result_BaselineMILP_ResiliencyPerformance_Size}
\end{figure}

Since the refrigerator is the most critical load for post-hurricane resiliency, we define \emph{resiliency performance} as the average time in hours per day for which primary load is serviced. Fig~\ref{fig:Result_BaselineMILP_ResiliencyPerformance_Size} shows the average time per day for which primary load was not serviced, for different system sizes when using the baseline controller. There is only one data point for the proposed controller in the plot; it corresponds to system size A, the smallest among the six studied. 
The figure shows that for the baseline controller to achieve a resiliency performance similar to what the proposed controller achieved with system A, the system cost/size has to be doubled. In other words, the equipment cost required to achieve a similar level of primary resiliency performance can be halved by the proposed controller.

\section{Conclusion}\label{section:Conclusions}

We present a novel MPC-based control system for providing resiliency during post-hurricane grid outages to a house with rooftop PV and a battery energy storage system. Simulation results show that the proposed controller significantly outperforms a baseline controller in servicing the primary load (refrigerator), and slightly outperforms in servicing the secondary loads (fans and lights). A comparison of resiliency provided by PV-battery system of different sizes with the baseline controller shows that the baseline controller requires a system size that is twice that of the proposed controller to provide the same level of resiliency performance. This study provides support to our premise that the cost of energy resiliency to loss of grid supply during nature disasters can be reduced significantly by using intelligent and automated decision making.

This study is a preliminary work which opens up many directions for future research. These include analysis of sensitivity to forecast errors, reducing the information requirements (both sensing and forecasts) of the control algorithm, optimal sizing of a PV-battery system taking into account resiliency during disasters and energy savings during normal times, and many more. 

\bibliographystyle{IEEEtran}
\bibliography{\DiCEbibPATH/resiliency,\DiCEbibPATH/disaster,\DiCEbibPATH/Barooah,\DiCEbibPATH/optimization,\DiCEbibPATH/systemid,\DiCEbibPATH/grid,\DiCEbibPATH/building,\DiCEbibPATH/systemid}

\begin{thebibliography}{10}
\providecommand{\url}[1]{#1}
\csname url@samestyle\endcsname
\providecommand{\newblock}{\relax}
\providecommand{\bibinfo}[2]{#2}
\providecommand{\BIBentrySTDinterwordspacing}{\spaceskip=0pt\relax}
\providecommand{\BIBentryALTinterwordstretchfactor}{4}
\providecommand{\BIBentryALTinterwordspacing}{\spaceskip=\fontdimen2\font plus
\BIBentryALTinterwordstretchfactor\fontdimen3\font minus
  \fontdimen4\font\relax}
\providecommand{\BIBforeignlanguage}[2]{{%
\expandafter\ifx\csname l@#1\endcsname\relax
\typeout{** WARNING: IEEEtran.bst: No hyphenation pattern has been}%
\typeout{** loaded for the language `#1'. Using the pattern for}%
\typeout{** the default language instead.}%
\else
\language=\csname l@#1\endcsname
\fi
#2}}
\providecommand{\BIBdecl}{\relax}
\BIBdecl

\bibitem{USGCRPreport:2014}
{U.S. Global Change Research Program}, ``2014: Highlights of climate change
  impacts in the {U}nited {S}tates: The third national climate assessment,''
  May 2014.

\bibitem{IrmaImpactReport:EIA:2017}
{United States Energy Information Agency}, ``{EIA-930 Hurricane {Irma} Impact
  Tracking Report Friday September 15, 2017, 15:00 hours},''
  \url{https://www.eia.gov/special/disruptions/archive/hurricane/irma/pdf/Irma_20170915_1500_report.pdf},
  2017.

\bibitem{GallucciRebuilding:spectrum:2018}
M.~Gallucci, ``{Rebuilding Puerto Rico's Power Grid: The Inside Story},''
  \emph{{IEEE Spectrum}}, 2018.

\bibitem{KishoreMortality:NEJM:2018}
N.~Kishore, D.~Marqu\'{e}s, A.~Mahmud, M.~V. Kiang, I.~Rodriguez, A.~Fuller,
  P.~Ebner, C.~Sorensen, F.~Racy, J.~Lemery, L.~Maas, J.~Leaning, R.~A.
  Irizarry, S.~Balsari, and C.~O. Buckee, ``Mortality in {P}uerto {R}ico after
  hurricane {M}aria,'' \emph{New England Journal of Medicine}, vol. 379, no.~2,
  pp. 162--170, 2018.

\bibitem{EIAAnnual:2019}
{United States Energy Information Agency}, ``Annual electric power industry
  report (survey form no. eia-861),'' \emph{US Energy Information
  Administration, Washington, DC, USA}, 2019.

\bibitem{CdcKeep:2019}
{Centers for disease control and prevention}, ``Keep food and water safe after
  a disaster or emergency,''
  \url{https://www.cdc.gov/disasters/foodwater/facts.html}, 2019, last
  accessed: Jan, 17, 2020.

\bibitem{NorrisHow:2017}
J.~Norris, ``How to prepare your kitchen for a hurricane,''
  \url{http://floridafoodandfarm.com/featured/how-to-prepare-your-kitchen-for-a-hurricane/},
  2017, last accessed: Jan, 17, 2020.

\bibitem{OttensteinGet:2019}
M.~Ottenstein, ``Get organized: {U}se our hurricane checklist to prepare for
  this year's hurricane season,'' \url{https://bit.ly/2FW1mr7}, 2019, last
  accessed: Jan, 17, 2020.

\bibitem{DiEvent:2012}
A.~Di~Giorgio and L.~Pimpinella, ``An event driven smart home controller
  enabling consumer economic saving and automated demand side management,''
  \emph{Applied Energy}, vol.~96, pp. 92--103, 2012.

\bibitem{AnvariOptimal:2014}
A.~Anvari-Moghaddam, H.~Monsef, and A.~Rahimi-Kian, ``Optimal smart home energy
  management considering energy saving and a comfortable lifestyle,''
  \emph{IEEE Transactions on Smart Grid}, vol.~6, no.~1, pp. 324--332, 2014.

\bibitem{BrahmanOptimal:2015}
F.~Brahman, M.~Honarmand, and S.~Jadid, ``Optimal electrical and thermal energy
  management of a residential energy hub, integrating demand response and
  energy storage system,'' \emph{Energy and Buildings}, vol.~90, pp. 65--75,
  2015.

\bibitem{MarzbandOptimal:2017}
M.~Marzband, H.~Alavi, S.~S. Ghazimirsaeid, H.~Uppal, and T.~Fernando,
  ``Optimal energy management system based on stochastic approach for a home
  microgrid with integrated responsive load demand and energy storage,''
  \emph{Sustainable cities and society}, vol.~28, pp. 256--264, 2017.

\bibitem{SanjariAnalytical:2017}
M.~J. Sanjari, H.~Karami, and H.~B. Gooi, ``Analytical rule-based approach to
  online optimal control of smart residential energy system,'' \emph{IEEE
  Transactions on Industrial Informatics}, vol.~13, no.~4, pp. 1586--1597,
  2017.

\bibitem{PrinceResilience:2019}
J.-J.~P. A., P.~Haessig, R.~Bourdais, and H.~Gueguen, ``Resilience in energy
  management system: {A} study case,'' \emph{IFAC-PapersOnLine}, vol.~52,
  no.~4, pp. 395 -- 400, 2019, iFAC Workshop on Control of Smart Grid and
  Renewable Energy Systems CSGRES 2019.

\bibitem{MastersRenewable:2013}
G.~M. Masters, \emph{Renewable and efficient electric power systems}.\hskip 1em
  plus 0.5em minus 0.4em\relax John Wiley \& Sons, 2013.

\bibitem{TanakaOptimal:2012}
K.~Tanaka, A.~Yoza, K.~Ogimi, A.~Yona, T.~Senjyu, T.~Funabashi, and C.-H. Kim,
  ``Optimal operation of dc smart house system by controllable loads based on
  smart grid topology,'' \emph{Renewable Energy}, vol.~39, no.~1, pp. 132--139,
  2012.

\bibitem{FaimanAssessing:2008}
D.~Faiman, ``Assessing the outdoor operating temperature of photovoltaic
  modules,'' \emph{Progress in Photovoltaics: Research and Applications},
  vol.~16, no.~4, pp. 307--315, 2008.

\bibitem{CostanzoGrey:2013}
G.~T. {Costanzo}, F.~{Sossan}, M.~{Marinelli}, P.~{Bacher}, and H.~{Madsen},
  ``Grey-box modeling for system identification of household refrigerators: A
  step toward smart appliances,'' in \emph{2013 4th International Youth
  Conference on Energy (IYCE)}, June 2013, pp. 1--5.

\bibitem{CuiHybrid:2019}
B.~Cui, C.~Fan, J.~Munk, N.~Mao, F.~Xiao, J.~Dong, and T.~Kuruganti, ``A hybrid
  building thermal modeling approach for predicting temperatures in typical,
  detached, two-story houses,'' \emph{Applied Energy}, vol. 236, pp. 101 --
  116, 2019.

\bibitem{DioufThe:2019}
B.~Diouf and C.~Avis, ``The potential of li-ion batteries in ecowas solar home
  systems,'' \emph{Journal of Energy Storage}, vol.~22, pp. 295 -- 301, 2019.

\bibitem{gurobi:2019}
\BIBentryALTinterwordspacing
{Gurobi Optimization, {LLC}}, ``Gurobi optimizer reference manual,'' 2019.
  [Online]. Available: \url{http://www.gurobi.com}
\BIBentrySTDinterwordspacing

\end{thebibliography}

\end{document}